\documentclass[10pt]{IEEEtran}%
\usepackage{amsmath,amsfonts,amssymb,graphicx,algorithm,color,cite}
\newtheorem{theorem}{Theorem}

\newtheorem{definition}{Definition}

\begin{document}
\textwidth 7.2in \textheight 9.4in \topmargin -0.4in

\oddsidemargin -0.45in \evensidemargin -0.2in

\title{Auction-Based Distributed Resource Allocation for Cooperation
Transmission in Wireless Networks}
\author{Jianwei Huang$^{1,2}$, Zhu Han$^{3}$, Mung Chiang$^{1}$, and H. Vincent Poor$^{1}$\\
$^{1}$Department of Information Engineering, the Chinese University of Hong Kong, Sha
Tin, NT, Hong Kong.
\\$^{2}$Department of Electrical Engineering, Princeton University, Princeton, NJ,
USA.\\$^{3}$Electrical and Computer Engineering Department, Boise
State University, Boise, ID, USA \thanks{This research was
supported by the NSF Grants ANI-03-38807, CCR-02-05214,
CNS-06-25637, CCF-0448012, and CCF-0635034, an ONR YIP grant.}
\\\vspace{-30pt}  } \maketitle

\begin{abstract}
Cooperative transmission can greatly improve communication system
performance by taking advantage of the broadcast nature of
wireless channels. Most previous work on resource allocation for
cooperation transmission is based on centralized control. In this
paper, we propose two share auction mechanisms, the SNR auction
and the power auction, to distributively coordinate the resource
allocation among users. We prove the existence, uniqueness and
effectiveness of the auction results. In particular, the SNR
auction leads to a fair resource allocation among users, and the
power auction achieves a solution that is close to the efficient
allocation.
\end{abstract}


\section{Introduction}

The cooperative communication concept has recently been proposed
\cite{sendonaris1,Wornell3} as a means to take advantage of the
 broadcast nature of wireless channels by using relays as virtual
antennas to provide the advantages of  multiple input
multiple output (MIMO) transmission. Various cooperative protocols such as
amplify-and-forward, decode-and-forward, and estimate-and-forward have been
proposed (e.g., \cite{sendonaris1,Wornell3,Cover,EF}). The work in
\cite{Madsen} analyzes  cooperative schemes involving
dirty paper coding, while energy-efficient transmission is considered for
broadcast networks in \cite{Yates2}. In \cite{Luo}, the authors evaluate
cooperative-diversity performance when the best relay is chosen according to
the average SNR  as well as the outage probability of relay selection based on the
instantaneous SNR. In \cite{Bletsas}, the authors propose a distributed relay
selection scheme that requires limited network knowledge and is based on
instantaneous SNRs. In \cite{Han_WCNC}, relay section, power management, and
subcarrier assignment are investigated for multiuser OFDM networks.

In order to maximize the performance of the cooperative
transmission network, we need to consider the global channel
information, including those between source-destination,
source-relay, and relay-destination. Most existing work in this
area is based on centralized control, which requires considerable
overhead for signalling and measurement. In this paper, we focus
on designing \emph{distributed} resource allocation algorithms for
cooperative networks. In particular, we want to answer the
following two questions: 1) ``When to relay'', i.e., when is it
beneficial to use the relay? 2) ``How to relay'', i.e., how should
the relay allocate resources among multiple competing users?

We answer these two questions by designing an \emph{auction}-based
framework for cooperative resource allocation. Auctions have
recently been introduced into several areas of wireless
communications (e.g., time slot allocation \cite{SunZheMod03} and
power control \cite{dramitinos2004abr,HuangMonet05}).
%
This paper is closely related to the auction mechanisms proposed
in \cite{HuangMonet05}, where the authors considered distributed
interference management in a cognitive radio network without a
relay. In that case, a user can only obtain a positive
transmission rate when it obtains some shared system resource. The
problem considered here is significantly different due to the
existence of the relay and the possibility of achieving a positive
transmission rate without using the relay.

We consider two network objectives here: \emph{fairness} and
\emph{efficiency}. Both might be difficult to achieve even in a
centralized fashion. This is because users' rate increases are
non-smooth and non-concave in the relay's transmission power, and
thus the corresponding optimization problems are non-convex. We
propose two auction mechanism, the SNR auction and the power
auction, which achieve the desired network objectives in a
distributed fashion under suitable technical conditions. In both
auctions, each user decides ``when to relay'' based on a simple
threshold policy that is locally computable. The question of ``how
to relay'' is answered by a simple weighted proportional
allocation among users who use the relay. Simulation results show
that the power auction achieves an average of $95\%$ of the
maximum rate increase in a two-user network over a wide range of
relay locations. The SNR auction achieves a fair allocation among
users but leads to a much lower total rate increase. This reflects
a fairness-efficiency tradeoff that can be exploited by a system
designer.

This paper is organized as follows. The system model and network objectives are given in Section
\ref{sec:sys_mod}.
In Section \ref{Sec: Share Auction},  two share auction mechanisms are proposed, their
mathematical properties are analyzed, and mechanisms for achieving auction results in a distributed
fashion are shown. Simulation results are discussed in Section \ref{sec:simulation} and conclusions
are drawn in Section \ref{sec:conclusions}. Due to space limitations, all proofs are omitted in
this conference version of the paper.

\section{System Model and Network Objectives}

\label{sec:sys_mod}

\begin{figure}[tptb]
\centering
\includegraphics[width=2in]{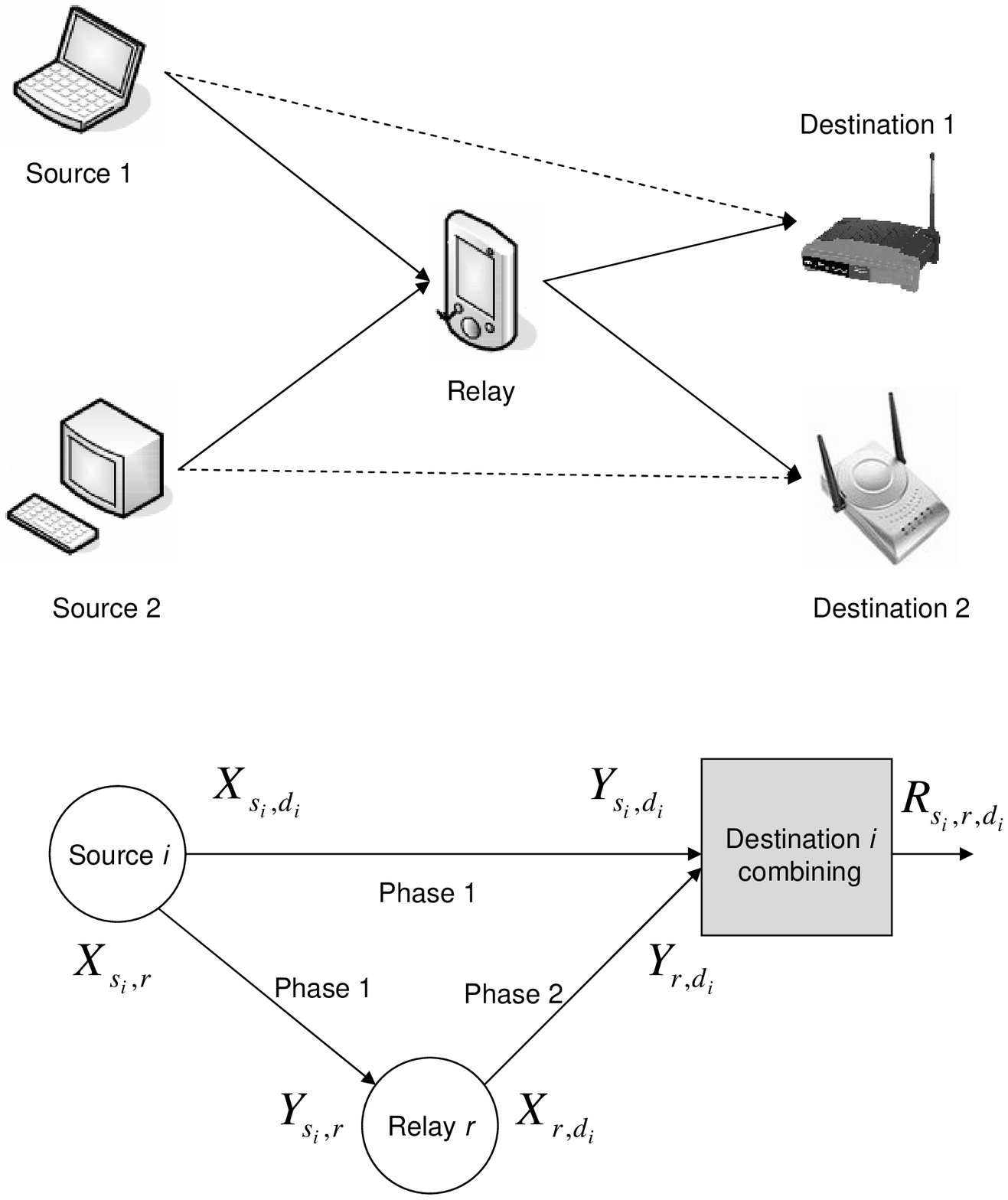} %
\caption{System model for cooperative transmission}%
\label{fig:system_model}%
\vspace{-5mm}\end{figure}

\subsection{System Model}

We focus our discussions on the amplify-and-forward (AF)
cooperative protocol \cite{Wornell3} in this paper. Other
cooperation protocols can be analyzed in a similar fashion. The
system diagrams are shown in Fig. \ref{fig:system_model}, where
there are one relay node $r$ and a set $\mathcal{I=}\left(
1,...,I\right)  $ of source-destination pairs. We also refer to
pair $i$ as \textit{user }$i$, which includes source node $s_{i}$
and destination node $d_{i}$.

For each user $i$, the cooperative transmission consists of two
phases. In Phase $1$, source $s_{i}$ broadcasts its information to
both destination $d_{i}$ and the relay $r$. The received signals
$Y_{s_{i},d_{i}}$ and $Y_{s_{i},r}$ at destination $d_{i}$ and
relay $r$ are given by
\begin{equation}
Y_{s_{i},d_{i}}=\sqrt{P_{s_{i}}G_{s_{i},d_{i}}}X_{s_{i}}+n_{d_{i}%
},\label{eqn:Ysd}%
\end{equation}
and
\begin{equation}
Y_{s_{i},r}=\sqrt{P_{s_{i}}G_{s_{i},r}}X_{s_{i}}+n_{r},\label{eqn:Ysri}%
\end{equation}
where $P_{s_{i}}$ represents the transmit power of source $s_{i}$, $X_{s_{i}}$
is the transmitted information symbol with unit energy at Phase $1$ at source
$s_{i}$, $G_{s_{i},d_{i}}$ and $G_{s_{i},r}$ are the channel gains from
$s_{i}$ to destination $d_{i}$ and relay $r,$ respectively, and $n_{d_{i}}$ and
$n_{r}$ are additive white Gaussian noises. Without loss of generality, we
assume that the noise level is the same for all of the links, and is denoted by
$\sigma^{2}$. We also assume that the channels are stable over each transmission frame.

The signal-to-noise ratio (SNR) at destination $d_{i}$ in Phase 1
is
\begin{equation}
\Gamma_{s_{i},d_{i}}=\frac{{P_{s_{i}}G_{s_{i},d_{i}}}}{{\sigma^{2}}%
}.\label{eqn:SNR_s_d}%
\end{equation}

For amplify-and-forward cooperative transmission, in Phase $2$ relay $r$
amplifies $Y_{s_{i},r}$ and forwards it to destination $d_{i}$ with
transmitted power $P_{r,d_{i}}$. The received signal at destination $d_{i}$
is
\begin{equation}
Y_{r,d_{i}}=\sqrt{P_{r,d_{i}}G_{r,d_{i}}}X_{r,d_{i}}+n_{d_{i}}^{\prime
},\label{eqn:relay to dest signal}%
\end{equation}
where
\begin{equation}
X_{r,d_{i}}=\frac{Y_{s_{i},r}}{|Y_{s_{i},r}|}\label{eqn:trans signal at ri}%
\end{equation}
is the unit-energy transmitted signal that relay $r$  receives from source $s_{i}$ in
Phase $1$, $G_{r,d_{i}}$ is the channel gain from relay $r$ to destination $d_{i}$, and
$n_{d_{i}}^{\prime}$ is the received noise at Phase $2$. Substituting (\ref{eqn:Ysri})
into (\ref{eqn:trans signal at ri}), we can rewrite (\ref{eqn:relay to dest signal}) as
\begin{equation}
Y_{r,d_{i}}=\frac{\sqrt{P_{r,d_{i}}G_{r,d_{i}}}(\sqrt{P_{s_{i}}G_{s_{i},r}%
}X_{s_{i},d_{i}}+n_{r})}{\sqrt{P_{s_{i}}G_{s_{i},r}+\sigma^{2}}}+n_{d_{i}%
}^{\prime}.\label{eqn:dest signal}%
\end{equation}
Using (\ref{eqn:dest signal}), the relayed SNR at destination $d_{i}$ with the help of
relay is
\begin{equation}
\Gamma_{s_{i},r,d_{i}}=\frac{P_{r,d_{i}}P_{s_{i}}G_{r,d_{i}}G_{s_{i},r}%
}{\sigma^{2}(P_{r,d_{i}}G_{r,d_{i}}+P_{s_{i}}G_{s_{i},r}+\sigma^{2}%
)}.\label{eqn:relay SNR}%
\end{equation}

If user $i$ performs only the direct transmission in Phase 1
(i.e., not using the relay), it achieves a total information rate
of
\begin{equation}
R_{s_{i},d_{i}}=W\log_{2}\left(  1+\Gamma_{s_{i},d_{i}}\right)
,\label{eqn:rate_s_d}%
\end{equation}
where $W$ is the signal bandwidth. On the other hand, if user $i$ performs the
transmissions in both Phases 1 and 2, it can then achieve a total information rate at the
output of maximal ratio combining as
\begin{equation}
R_{s_{i},r,d_{i}}=\frac{1}{2}W\log_{2}\left(  {1+\Gamma_{s_{i},d_{i}}+}%
\Gamma_{s_{i},r,d_{i}}\right)  .\label{eqn: MRC output}%
\end{equation}
The coefficient $1/2$ is used to model the fact that cooperative transmission
will occupy one out of two phases (e.g., time, bandwidth, code). Since
$\Gamma_{s_{i},r,d_{i}}$ is the extra SNR increase compared with the direct
transmission, we also denote
\begin{equation}
\bigtriangleup\mathtt{SNR}_{i}\triangleq\Gamma_{s_{i},r,d_{i}}.
\end{equation}

\begin{figure}[t]
\centering
\includegraphics[width=2in]{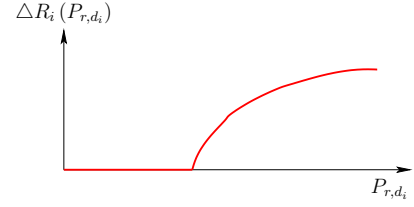} \caption{Rate increase
as a function of relay transmission power}%
\label{fig:rate-increase}%
\end{figure}

Based on (\ref{eqn:rate_s_d}) and (\ref{eqn: MRC output}), the rate increase
that user $i$ obtains by cooperative transmission is%
\begin{equation}
\bigtriangleup R_{i}=\max\left\{  R_{s_{i},r,d_{i}}-R_{s_{i},d_{i}},0\right\}
,\label{eq:rate_increase}%
\end{equation}
which is nonnegative since the source can always choose not to use the relay
and thereby obtain zero rate increase.
%
$\bigtriangleup R_{i}$ is a function of the channel gains of the source-destination, source-relay
and relay-destination links, as well as the transmission power of the source and the relay. In
particular, $\bigtriangleup R_{i}$ is a non-decreasing, non-smooth, and non-concave function of the
relay transmission power $P_{r,d_{i}}$, as illustrated in Fig.~\ref{fig:rate-increase}.

We assume that the source transmission power $P_{s_{i}}$ is fixed for each user $i$, as
well as the relay's total power, $P$. The relay determines the allocation of its
transmission power among users, $\boldsymbol{P}_{r}\triangleq\left(  P_{r,d_{1}%
},...,P_{r,d_{I}}\right)  $, such that the total power constraint is not
violated, i.e.,
\begin{equation}
\boldsymbol{P}_{r}\in\mathcal{P}_{r}\triangleq\left\{  \boldsymbol{P}%
_{r}\left\vert \sum_{i}P_{r,d_{i}}\leq P,P_{r,d_{i}}\geq0,\forall
i\in\mathcal{I}\right.  \right\}  .
\end{equation}

\subsection{Network Objectives: Efficiency and Fairness}

We consider two different network objectives: \emph{efficiency} and
\emph{fairness}. An efficient power allocation $\boldsymbol{P}_{r}%
^{\text{efficient}}$ maximizes the total rate increase of all users by solving
the following problem,
\begin{equation}
\max_{\boldsymbol{P}_{r}\in\mathcal{P}_{r}}\sum_{i\in\mathcal{I}%
}\bigtriangleup R_{i}\left(  P_{r,d_{i}}\right)  .\label{Problem:efficiency}%
\end{equation}

In many cases, an efficient allocation discriminates against users who are far away
from the relay. To avoid this, we also consider a fair power allocation
$\boldsymbol{P}_{r}^{\text{fair}}$, which solves the following problem%
\begin{align}
\min_{\boldsymbol{P}_{r}\in\mathcal{P}_{r}}\;\; & c\label{Problem:Fairness}\\
\mbox{subject to}\;\; &  \frac{\bigtriangleup R_{i}\left(  \bigtriangleup
\mathtt{SNR}_{i}\right)  }{\partial\left(  \bigtriangleup\mathtt{SNR}%
_{i}\right)  }=c\cdot \boldsymbol{1}_{\{\bigtriangleup \mathtt{SNR}_{i}>0\}},\forall
i\in\mathcal{I}.\nonumber
\end{align}
Here $\boldsymbol{1}_{\{\cdot\}}$ is the indicator function. The intuition behind Problem
(\ref{Problem:Fairness}) is that for all users that choose to use the relay, the
corresponding $\bigtriangleup \mathtt{SNR}$ should be maximized subject to the same
marginal utility among these users. This can be translated into the minimization of the
common marginal utility, due to the concavity of $\bigtriangleup R_{i}$ in terms of
$\bigtriangleup \mathtt{SNR}_{i}$ (within the appropriate region). As an example, when
the direct transmission SNR $\Gamma_{s_{i},d_{i}}$ is the same for all user $i$, the
constraint in Problem (\ref{Problem:Fairness}) means that
$\bigtriangleup\mathtt{SNR}_{i}$ is the same for all users with positive rate increase. A
numerical example of such fair allocation is shown in Section~\ref{sec:simulation}.

We notice that a fair allocation needs to be Pareto optimal, i.e., no user's rate can be
increased without decreasing the rate of another user. However, an efficient or fair
allocation need not fully utilize the resource at the relay, i.e.,
$\sum_{i\in\mathcal{I}}P_{r,d_{i}}$ can be less than $P$. This could happen, for example,
when the relay is far away from all users so that allowing the relay to transmit half of
the time will only decrease the total achievable rate. This is very different from most
previous network resource allocation problems (including \cite{HuangMonet05}), in which
the network performance is maximized only if the resource is fully utilized.

Since $\bigtriangleup R_{i}\left(  P_{r,d_{i}}\right)  $ is non-smooth and non-concave, it is well
known that Problems (\ref{Problem:efficiency}) and (\ref{Problem:Fairness}) are NP hard to solve
even in a centralized fashion. In the rest of the paper, we will propose two auction mechanisms
that can (approximately) solve these problems under suitable technical conditions in a distributed
fashion.

\section{Share Auction}

\label{Sec: Share Auction}

An auction is a decentralized market mechanism for allocating resources in an economy. An
auction consists of three key elements: 1) The \emph{good}, or the resource to be
allocated. 2) An \emph{auctioneer}, who determines the allocation of the good according
to the auction rules. 3) A group of \emph{bidders}, who want to obtain the good from the
auctioneer. The interactions and outcome of an auction are determined by the
\emph{rules}, which include four components: 1) The \emph{information} the auctioneer and
bidders know before the auction starts. 2) The \emph{bids} submitted to the auctioneer by
the bidders. 3) The \emph{allocation} determined by the auctioneer based on the bids. 4)
The \emph{payments} payed by the bidders to the auctioneer as functions of bids and
allocations.

In the cooperative network considered here, it is natural to design auction mechanisms in
which the \emph{good} is the relay's total transmit power $P$, the auctioneer is the
relay, and the bidders are the users. One well known auction mechanism that achieves the
efficient allocation is the VCG  auction \cite{Krishna2002}.
However, the VCG auction requires the relay to gather global
network information from the users, and solves $I+1$ nonconvex
optimization problems. This might be too complicated for real-time
implementations. To overcome the limitation of the VCG auction, we
propose two simpler share auctions, the \textit{SNR auction} and
the \textit{power auction}.\footnote{Both auctions are similar to
the ones proposed in \cite{HuangMonet05}. However, due to the
unique characteristics of the relay network, especially the
non-smooth and non-concave nature of the rate increase function
(e.g., Fig. \ref{fig:rate-increase}), the analysis is more
involved and the results are very different from those in
\cite{HuangMonet05}.} The main advantages of the two proposed
auctions in this section are the simplicities of bids and
allocation. The rules of the two auctions are described below,
with the only difference being in payment determination.

\textbf{\underline{Share Auction (SNR Auction and Power Auction)}}

\begin{itemize}
\item \emph{Information}: Besides the public and local information (i.e.,
$W,P,{\sigma^{2},P}_{s_{i}},G_{s_{i},d_{i}}$), each user $i$ also knows the
channel gains $G_{s_{i},r}$ and $G_{r,d_{i}}$, either through measurement or
explicit feedback from relay $r$. The relay announces a positive
\textit{reserve bid} $\beta>0$ and a \textit{price} $\pi>0$ to all users
before the auction starts.

\item \emph{Bids}: User $i$ submits bid $b_{i}\geq0$ to the relay.

\item \emph{Allocation}: The relay allocates transmit power according to
\begin{equation}
P_{r,d_{i}}=\frac{b_{i}}{\sum_{j\in\mathcal{I}}b_{j}+\beta}P.
\label{eq:power-allocation}%
\end{equation}

\item \emph{Payments}: In an SNR auction, source $i$ pays the relay $C_{i}%
=\pi\bigtriangleup\mathtt{SNR}_{i}$. In a power auction, source $i$ pays the
relay $C_{i}=\pi P_{r,d_{i}}$.
\end{itemize}


A bidding profile is defined as the vector containing the users' bids,
$\boldsymbol{b}=\left(  b_{1},...,b_{I}\right)  $. The bidding profile of user
$i$'s opponents is defined as $b_{-i}=\left(  b_{1},...,b_{i-1,}%
b_{i+1},...,b_{I}\right)  $, so that $\boldsymbol{b}=\left(  b_{i}%
;b_{-i}\right)  .$ User $i$ chooses $b_{i}$ to maximize its payoff
\begin{equation}
U_{i}\left(  b_{i};b_{-i},\pi\right)  =\bigtriangleup R_{i}\left(  P_{r,d_{i}%
}\left(  b_{i};b_{-i}\right)  \right)  -C_{i}\left(  b_{i};b_{-i},\pi\right)
.
\end{equation}
For notational simplicity, we omit the dependence on $\beta$ and other system parameters.

If the reserve bid $\beta=0$, then the resource allocation in
(\ref{eq:power-allocation}) depends only on the ratio of the bids.
A bidding profile $k\boldsymbol{b}$ (for any $k>0$) leads to the
same resource allocation as $\boldsymbol{b}$, which is not
desirable in practice. That is why we need a positive reserve bid.
However, the value of $\beta$ is not important as long as it is
positive. For example, if we increase $\beta$ to
$k^{\prime}\beta$, then users can just scale $\boldsymbol{b}$ to
$k^{\prime }\boldsymbol{b}$, which leads to the same resource
allocation. For simplicity, we will choose $\beta=1$ in all the
simulations in Section \ref{sec:simulation}.

The desirable outcome of an auction is called a \emph{Nash Equilibrium} (NE),
which is a bidding profile $\boldsymbol{b}^{\ast}$ such that no user wants to
deviate unilaterally, i.e.,
\begin{equation}
U_{i}\left(  b_{i}^{\ast};b_{-i}^{\ast},\pi\right)  \geq U_{i}\left(
b_{i};b_{-i}^{\ast},\pi\right)  ,\forall i\in\mathcal{I},\forall b_{i}\geq0.
\end{equation}
Define user $i$'s \textit{best response} (for fixed $b_{-i}$ and price $\pi$)
as
\begin{equation}
\mathcal{B}_{i}\left(  b_{-i},\pi\right)  =\left\{  b_{i}\left\vert b_{i}%
=\arg\max_{\tilde{b}_{i}\geq0}U_{i}\left(  \tilde{b}_{i};b_{-i},\pi\right)
\right.  \right\}  ,\label{eq:best-response}%
\end{equation}
which in general could be a set. An NE is also a fixed point
solution of all users' best responses. We would like to answer the
following four questions for both auctions: 1) When does an NE
exist? 2) When is the NE unique? 3) What are the properties of the
NE? 4) How can the NE be reached in a distributed fashion?

\subsection{SNR Auction}

Let us first determine the users' best responses (e.g., (\ref{eq:best-response}%
)) in the SNR auction, which clearly depend on the price $\pi$. For each user
$i$, there are two critical price values, $\underline{\pi}_{i}^{s}$ and
$\hat{\pi}_{i}^{s}$, where%
\begin{equation}
\underline{\pi}_{i}^{s}\triangleq\frac{W}{2\ln2\left(  1+{\Gamma_{s_{i},d_{i}%
}+}\frac{PG_{r,d_{i}}P_{s_{i}}G_{s_{i},r}}{\left(  P_{s_{i}}G_{s_{i}%
,r}+PG_{r,d_{i}}+\sigma^{2}\right)  \sigma^{2}}\right)  },
\end{equation}
and $\hat{\pi}_{i}^{s}$ is the smallest positive root of
\begin{multline}
g_{i}^{s}\left(  \pi\right)  \triangleq\pi\left(  1+{\Gamma_{s_{i},d_{i}}%
}\right) \\
-\frac{W}{2}\left(  \log_{2}\left(  \frac{2\pi\ln2}{W}\left(  1+{\Gamma
_{s_{i},d_{i}}}\right)  ^{2}\right)  +\frac{1}{\ln2}\right).  \label{eq:gi}%
\end{multline}
Both $\underline{\pi}_{i}^{s}$ and $\hat{\pi}_{i}^{s}$ can be calculated locally by user
$i$. 

\begin{theorem}
\label{Theorem:SNR_Auction_BestResponse}In an SNR auction, user $i$'s unique best
response function is%
\begin{equation}
\mathcal{B}_{i}\left(  b_{-i},\pi\right)  =f_{i}^{s}\left(  \pi\right)
\left(  b_{-i}+\beta\right)  .\label{eq:SNR_best_response}%
\end{equation}
If $\hat{\pi}_{i}^{s}>\underline{\pi}_{i}^{s}$, then
\begin{multline}
f_{i}^{s}\left(  \pi\right)  =\label{eq:SNR_auction_BR}\\
\left\{
\begin{array}
[c]{cc}%
\infty, & \pi\leq\underline{\pi}_{i}^{s}\\
\text{{\small $\frac{\left(  P_{s_{i}}G_{s_{i},r}+\sigma^{2}\right)
\sigma^{2}}{\frac{PG_{r,d_{i}}P_{s_{i}}G_{s_{i},r}}{\frac{W}{2\pi\ln
2}-1-{\Gamma_{s_{i},d_{i}}}}-\left(  P_{s_{i}}G_{s_{i},r}+PG_{r,d_{i}}%
+\sigma^{2}\right)  \sigma^{2}},$}} & \pi\in\left(  \underline{\pi}_{i}%
^{s},\hat{\pi}_{i}^{s}\right)  \\
0, & \pi\geq\hat{\pi}_{i}^{s}%
\end{array}
\right.  .
\end{multline}
If $\hat{\pi}_{i}^{s}<\underline{\pi}_{i}^{s}$, then $f_{i}^{s}\left(
\pi\right)  =\infty$ for $\pi<\hat{\pi}_{i}^{s}$ and $f_{i}^{s}\left(
\pi\right)  =0$ for $\pi\geq\hat{\pi}_{i}^{s}$.
\end{theorem}

\begin{figure}[ptb] \centering
\includegraphics[width=1.6in]{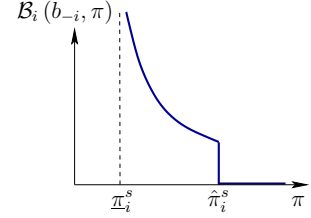}\caption{User $i$'s best
response in the SNR auction if $\underline{\pi}_{i}^{s}<\hat{\pi}_{i}^{s}$.}%
\label{fig:SNR-br}%
\vspace{-5mm}\end{figure}

First consider the case in which $\hat{\pi}_{i}^{s}>
\underline{\pi}_{i}^{s}$, where $\mathcal{B}_{i}\left(
b_{-i},\pi\right)  $ is illustrated in Fig.~\ref{fig:SNR-br}. The
price $\hat{\pi}_{i}^{s}$ determines when it is beneficial for
user $i$ to use the relay.\ With any price larger than
$\hat{\pi}_{i}^{s}$, user $i$ cannot obtain a positive payoff from
the auction no matter what bid it submits, and thus it should
simply use direct transmission and achieve a rate of
$R_{s_{i},d_{i}}$. As a result, $\mathcal{B}_{i}\left(
b_{-i},\pi\right)  $ is discontinuous at $\hat{\pi}_{i}^{s}$. When
$\pi \in\left( \underline{\pi}_{i}^{s},\hat{\pi}_{i}^{s}\right) $,
user $i$ wants to participate in the auction, and its best
response depends how much other users bid ($b_{-i}$). When the
price is smaller than $\underline{\pi}_{i}^{s}$, user $i$ becomes
so aggressive that it demands a large SNR increase that cannot be
achieved even of all the resource is allocated to it. This is
reflected by an infinite bid in (\ref{eq:SNR_auction_BR}). Now
consider the case in which
$\hat{\pi}_{i}^{s}<\underline{\pi}_{i}^{s}$. User $i$ either
cannot obtain a positive payoff or cannot achieve the desired SNR\
increase, and thus the best response is either
$0$ or $\infty$.%

Combining (\ref{eq:power-allocation}) and (\ref{eq:SNR_auction_BR}), we know that if an NE\ exist,
the relay power allocated for user $i$ is
\begin{equation}
P_{r,d_{i}}\left(  \pi\right)  =\frac{f_{i}^{s}\left(  \pi\right)  }{f_{i}%
^{s}\left(  \pi\right)  +1}P,
\end{equation}
and $\sum_{i\in\mathcal{I}}\frac{f_{i}^{s}\left(  \pi\right)  }{f_{i}%
^{s}\left(  \pi\right)  +1}<1$. The strictly inequality is due to the positive
reserve bid $\beta$.

Next we need to find the fixed point of all users' best responses, i.e., the
NE. A trivial case would be $\hat{\pi}_{i}^{s}\leq\underline{\pi}_{i}^{s}$ for
all user $i$, in which case there exists a unique all-zero NE $\boldsymbol{b}%
^{\ast}=\boldsymbol{0}$. The more interesting case would be the following.

\begin{definition}
A network is \emph{SNR-regular} if there exists at least one user $i$ such
that $\hat{\pi}_{i}^{s}>\underline{\pi}_{i}^{s}$.
\end{definition}

\begin{theorem}
\label{Theorem:SNR_auction}Consider an SNR auction in an SNR-regular network.
There exists a threshold price $\pi_{th}^{s}$ such that a unique NE exists if
$\pi>\pi_{th}^{s}$; otherwise no NE exists.
\end{theorem}

Unlike the result in \cite{HuangMonet05}, the unique NE in Theorem
\ref{Theorem:SNR_auction} might not be a continuous function of
$\pi$, due to the discontinuity of the best response function as
shown in Fig. \ref{fig:SNR-br}. This has been observed in the
simulation results described in Section \ref{sec:simulation}. In
particular, the unique NE could be all zero for any price
$\pi>\pi_{th}^{s}$, even if the network is SNR-regular.

It can be seen that the \textquotedblleft marginal utility
equalization\textquotedblright\ property of a fair allocation (i.e., the
constraint in Problem (\ref{Problem:Fairness})) is satisfied at the NE of the
SNR auction. However, there always exists some \textquotedblleft resource
waste\textquotedblright\ since some power will never be allocated to any user
because of the positive reserve bid $\beta$. However, by choosing a price
$\pi$ larger than, but very close to, $\pi_{th}^{s}$, we could reduce the
resource waste to a minimum and approximate the fair allocation. Formally, we
define a reduced feasible set parameterized by $\delta$ as%
\begin{equation}
\mathcal{P}_{r}^{\delta}\triangleq\left\{  \boldsymbol{P}_{r}\left\vert
\sum_{i}P_{r,d_{i}}\leq P\left(  1-\delta\right)  ,P_{r,d_{i}}\geq0,\forall
i\in\mathcal{I}\right.  \right\}  .
\end{equation}
Then we can show the following.

\begin{theorem}
\label{Theorem:SNR_auction_fairness}Consider an SNR auction in an SNR-regular
network, where $f_{i}^{s}\left(  \pi\right)  $ is continuous at $\pi_{th}^{s}$
for each user $i$, and greater than zero for at least one user. For any sufficiently small $\delta$,
there exists a price $\pi^{s,\delta}$ under which
the unique NE\ achieves the fair allocation $\boldsymbol{P}_{r}^{\text{fair}}$
with a reduced feasible set $\mathcal{P}_{r}^{\delta}$.
\end{theorem}

A sufficiently small $\delta$ makes sure that we deal with a regime in which
$f_{i}^{s}\left( \pi\right)  $ is continuous for any user $i$. This is also desirable in
practice since we want to minimize the amount of resource wasted.

\subsection{Power Auction}

The best response function in the power auction is nonlinear and complicated in
general. However, in the special case of low SNR where ${\Gamma_{s_{i},d_{i}}%
}$ and $\bigtriangleup\mathtt{SNR}_{i}\left(  b_{i},b_{-i}\right)  $ are small
for all $i$, i.e.,
\begin{multline}
W\log_{2}\left(  1+{\Gamma_{s_{i},d_{i}}+}\bigtriangleup\mathtt{SNR}%
_{i}\left(  b_{i},b_{-i}\right)  \right)  \\
\approx\frac{W}{\ln2}\left(  {\Gamma_{s_{i},d_{i}}+}\bigtriangleup
\mathtt{SNR}_{i}\left(  b_{i},b_{-i}\right)  \right)  ,
\end{multline}
$\mathcal{B}_{i}\left(  b_{-i},\pi\right)  $ has a linear form similar to that in
(\ref{eq:SNR_auction_BR}). For each user, we can similarly define $f_{i}%
^{p}\left(  \pi\right)  $, $\underline{\pi}_{i}^{p}$,
$\hat{\pi}_{i}^{p}$ and $g_{i}^{p}\left(  \pi\right)  $ as in the
SNR\ auction case. One key difference here is that the value of
$\hat{\pi}_{i}^{p}$ depends on the relative relationship between
$G_{s_{i},d_{i}}$ and $G_{s_{i},r}$. If
$G_{s_{i},d_{i}}>G_{s_{i},r}$, then $\hat{\pi}_{i}^{p}=0$ and user
$i$ never uses the relay. If $G_{s_{i},d_{i}}<G_{s_{i},r}$, then
$\hat{\pi}_{i}^{p}$ is the smallest positive root of
$g_{i}^{p}\left(  \pi\right)  $. Details are omitted due to space
limitations.

In terms of the existence, uniqueness and properties of the NE, we have the following.

\begin{definition}
A network is \textit{power-regular} if $\hat{\pi}_{i}^{p}>\underline{\pi}%
_{i}^{p}$ for at least one user $i$.
\end{definition}

\begin{theorem}
\label{Theorem:Power_auction}Consider a power auction in a power-regular
network with low SNR. There exists a threshold price $\pi_{th}^{p}>0$ such that
a unique NE exists if $\pi>\pi_{th}^{p}$; otherwise no NE\ exists.
\end{theorem}

\begin{theorem}
Consider a power auction in a power-regular network with low SNR, where $f_{i}^{p}\left( \pi\right)
$ is continuous at $\pi_{th}^{p}$ for each user $i$, and greater than zero for at least one user.
For any sufficiently small $\delta$, there exists a price $\pi^{p,\delta}$ under which the unique
NE\ achieves the efficient allocation $\boldsymbol{P}_{r}^{\text{efficient}}$ with a reduced
feasible set $\mathcal{P}_{r}^{\delta}$.
\end{theorem}

\subsection{Distributed Iterative Best Response Updates}

\label{sect:distributed_algorithm}

The last question we want to answer is how the NE can be reached in a distributed
fashion. Consider the SNR auction as an example. It is clear that the best response
function in (\ref{eq:SNR_auction_BR}) can be calculated in a distributed fashion with
limited information feedback from the relay. However, each user does not have enough
information to calculate the best response of other users, which prevents it from
directly calculating the NE. Nevertheless, the NE can be achieved in a distributed
fashion if we allow the users to \emph{iteratively} submit their bids based on best
response functions.

Suppose users update their bids $\boldsymbol{b}\left(  t\right)  $ at time $t$ according
to the best response functions as in (\ref{eq:SNR_best_response}), based on other users'
bids $\boldsymbol{b}\left(  t-1\right)  $ in the previous time $t-1$, i.e.,
\begin{equation}
\boldsymbol{b}\left(  t\right)  =\boldsymbol{F}^{s}\left(  \pi\right)
\boldsymbol{b}\left(  t-1\right)  +\boldsymbol{f}^{s}\left(  \pi\right)
\beta\text{,}\label{eq:SNR_best_response_udpates}%
\end{equation}
where both $\boldsymbol{b}\left(  t\right)  $ and
$\boldsymbol{b}\left( t-1\right)  $ are column vectors,
$\boldsymbol{F}^{s}\left(  \pi\right)  $ is an $I$-by-$I\ $ matrix
whose $\left(  i,j\right)  $th component equals $f_{i}^{s}\left(
\pi\right)  $, and $\boldsymbol{f}^{s}\left(  \pi\right) =\left[
f_{1}^{s}\left(  \pi\right)  ,...,f_{I}^{s}\left(  \pi\right)
\right]  ^{\prime}$.

\begin{theorem}
\label{Theorem:convergence} If there exists a unique nonzero NE in the SNR auction, the
best response updates in (\ref{eq:SNR_best_response_udpates})  globally and geometrically
converge to the NE from any positive $\boldsymbol{b}\left(  0\right)  $.
\end{theorem}

Similar convergence results can be proved for the power auction.


\vspace{-10pt}
\section{Simulation Results}

\label{sec:simulation}

\begin{figure}[ptb]
\centering
\includegraphics[width=2in]{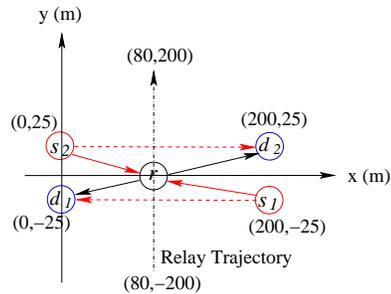} \caption{A two-user cooperative
network}%
\label{fig:2user}%
\vspace{-5mm}\end{figure}

We first simulate various auction mechanisms for a two-user network. As shown
in Fig.~\ref{fig:2user}, the locations of the two sources ($s_{1}$ and $s_{2}%
$) and two destinations ($d_{1}$ and $d_{2}$) are fixed at
(200m,-25m), (0m,25m), (0m,-25m), and (200m,25m). We fix the $x$
coordinate of the relay node $r$ at 80m and its $y$ coordinate
varies within the range [-200m,200m]. In the simulation, the relay
moves along a line. The propagation loss factor is set to 4, and
the channel gains are distance based (i.e., time-varying fading is
not considered here). The transmit power between a source and its
destination is $P_{s_{i}}=0.01$W for all user $i$, the noise level
is $\sigma^{2}=10^{-11}$W, and the bandwidth is $W=1$MHz. The
total power of the relay node is set to $P=0.1$W.

\begin{figure}[tb]
\centering
\includegraphics[width=2.2in]{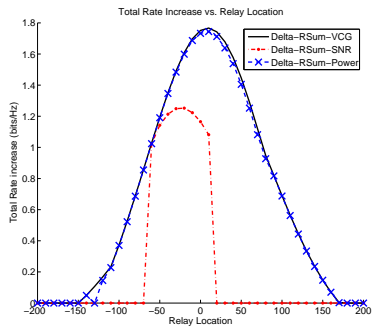} \caption{Total rate increases
vs. relay location (y-axis) for three auctions.}%
\label{fig:2user-total}%
\vspace{-3mm}\end{figure}

\begin{figure}[tb]
\centering
\includegraphics[width=2.2in]{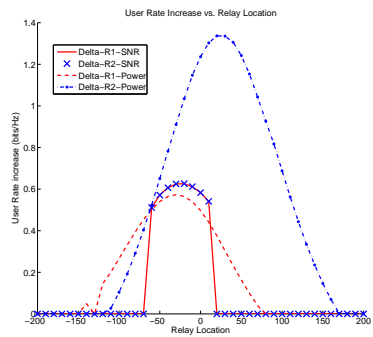} \caption{Individual rate
increases vs. relay location (y-axis) for the SNR auction and the power auction.}%
\label{fig:2user-indivisual}%
\vspace{-5mm}\end{figure}

In Fig.~\ref{fig:2user-total}, we show the total rate increases
achieved by two users in three auctions. The VCG auction achieves
the efficient allocation by solving three non-convex optimization
problems by the relay. For both the SNR auction and the power
auction, the resource allocation depends on the choice of price
$\pi$ (but is independent of the reserve bid $\beta$). Every point
on the curve represents an allocation in which the price is
adjusted so that the total resource allocated to both users is
more than $0.99P$ (unless this is not possible). The power auction
achieves performance very close to that of the VCG auction. At
those locations where the VCG auction achieves a positive rate
increase, the power auction achieves a rate increase with an
average of $95\%$ of that achieved by the VCG auction. The SNR
auction achieves less total rate increases but leads to fair
resource allocations when both users use the relay (as can be seen
in Fig.~\ref{fig:2user-indivisual}).

In Fig.~\ref{fig:2user-indivisual}, we show the individual rate
increases of both users in the SNR auction and  the power auction.
The individual rate increases in the VCG auction are similar to
that of the power auction and thus are not shown here. First
consider the power auction. Since the relay movement trajectory is
relatively closer to source $s_{2}$ than to source $s_{1}$, user
$2$ achieves an overall better performance compared with user $1$.
In particular, user $2$ achieves a peak rate increase of $1.35$
bits/Hz when the relay is at location 25m ($y$-axis), compared
with the peak rate increase of $0.56$ bits/Hz achieved by user $1$
when relay is at location -25m. Things are very different in an
SNR auction, where the resource allocation is fair. In particular,
since the distance between a source and its destination is the
same for both users in our simulation, both users achieve the same
positive rate increases when they both use the relay. This is the
case when the relay is between locations -60m and 10m. At other
locations, users just choose not to use the relay since they
cannot both get equal rate increases while obtaining a positive
payoff. This shows the tradeoff between efficiency and fairness.

\begin{figure}[t]
\centering
\includegraphics[width=2.2in]{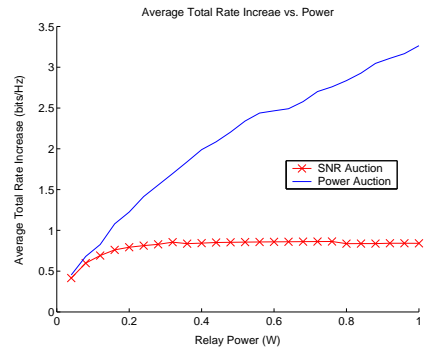} \caption{Total network
rate increase vs. relay power.}%
\vspace{-3mm}\label{fig:multi-user-rate}%
\end{figure}

\begin{figure}[t]
\centering
\includegraphics[width=2.2in]{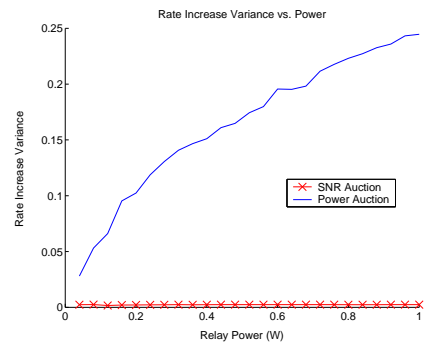} \caption{Total
positive rate increase variance vs. relay power.}%
\label{fig:multi-user-variance}%
\vspace{-5mm}\end{figure}

Next, we consider the case in which there are multiple users in
the network. To be specific, there are 20 users in the network,
with their source nodes and destination nodes randomly and
uniformly located within the square field that has the same range
of [-150m,150m] on both the $x$-axis and the $y$-axis. A single
relay is fixed at the location  (0m,0m). The total transmission
power $P$ of the relay is varied between $0.04$ W and $1$ W.
Figs.~\ref{fig:multi-user-rate} and \ref{fig:multi-user-variance}
show the corresponding simulation results. Each point in the
figures represents results averaged over $100$ randomly generated
network topologies. With an increasing amount of resource at the
relay node, the total network rate increase improves in both
auctions (as seen in Fig.~\ref{fig:multi-user-rate}), and the
power auction achieves higher rate increase than the SNR auction.
Fig.~\ref{fig:multi-user-variance} shows the variance of the rate
increase (among the users with positive rate increase), and it is
clear that the SNR auction achieves a fair resource allocation as
indicated by the almost zero variance in all cases.


\section{Conclusions}
\label{sec:conclusions}

Cooperative transmission can greatly improve communication system
performance by taking advantage of the broadcast nature of
wireless channels and cooperation among users. In this paper, we
have proposed two share auction mechanisms, the SNR auction and
the power auction, to distributively coordinate the relay power
allocation among users. We have proven the existence and
uniqueness of the Nash equilibrium in both auctions.  Under
suitable conditions, the SNR auction achieves the fair allocation,
while the power auction achieves the efficient allocation.
Simulations results for both two-user and multiple-user networks
have been used to demonstrate the effectiveness of the auction
mechanisms. In particular, the power auction achieves an average
of $95\%$ of the maximum rate in the two-user case under a wide
range of relay locations, and the SNR auction leads to a
performance improvement having small variation among users.

\end{document}